\documentclass[12pt,letterpaper]{JHEP3}         
\usepackage{amssymb,amsfonts}

\usepackage{epsfig}

\def\be{\begin{eqnarray}}
\def\ee{\end{eqnarray}}
\newcommand{\nn}{\nonumber}
\newcommand\para{\paragraph{}}

\newcommand{\eqn}[1]{(\ref{#1})}

\def\Dslash{\,\,{\raise.15ex\hbox{/}\mkern-12mu D}}
\def\Dbarslash{\,\,{\raise.15ex\hbox{/}\mkern-12mu {\bar D}}}
\def\delslash{\,\,{\raise.15ex\hbox{/}\mkern-9mu \partial}}
\def\delbarslash{\,\,{\raise.15ex\hbox{/}\mkern-9mu {\bar\partial}}}
\def\pslash{\,\,{\raise.15ex\hbox{/}\mkern-9mu p}}
\def\calDslash{\,\,{\raise.15ex\hbox{/}\mkern-12mu {\cal D}}}

\def\lae{\mathrel{\mathop{\smash{\lower .5 ex \hbox{$\stackrel<\sim$}}}}}
\def\lae{\mathrel{\mathop{\smash{\lower .5 ex \hbox{$\stackrel>\sim$}}}}}

\title{Fluctuation and Dissipation at a Quantum Critical Point}

\author{David Tong and Kenny Wong  \\

Department of Applied Mathematics and Theoretical Physics, University of Cambridge, UK \\ 

{\tt d.tong, k.wong@damtp.cam.ac.uk}
}

\abstract{In non-relativistic field theories,  quantum fluctuations give rise to dissipative behaviour even at zero temperature.  Here we use holographic methods to explore the
dissipative dynamics of massive particles coupled to quantum critical theories. We present analytic expressions for
correlation functions and response functions. The behaviour changes qualitatively as the dynamical exponent passes through $z=2$. In particular,
for $z>2$, the long time dynamics of the particle is independent of its inertial mass.}

\begin{document}
\pagestyle{plain} \setcounter{page}{1}
\newcounter{bean}
\baselineskip16pt

\section{Introduction}

At a  {\it quantum critical point}, physics is invariant under the dynamical scaling symmetry
\be \vec{x}\rightarrow \lambda \vec{x} \ \ \ \ ,\ \ \ \ t\rightarrow \lambda^z t\label{scaling}\ee
$z$ is called the {\it dynamical exponent}. Such theories have been studied in great detail, starting with \cite{hertz,millis}. For $z=1$, the theory is typically Lorentz invariant, with the scaling symmetry  promoted to the full conformal group. Theories with $z\neq 1$ arise naturally in a number of condensed matter contexts.

\para
A gravity dual for non-relativistic critical points was first proposed in \cite{shamit}. The  $(d+1)$-dimensional background metric is known as the {\it Lifshitz geometry} and takes the simple form
\be ds^2 =  -r^{2z}dt^2+\frac{dr^2}{r^2} +r^2d\vec{x}^2\label{metric}\ee
where we have set the curvature of spacetime to unity. 
Several physical phenomena have been studied in this background, including mechanisms for linear DC conductivity and power-law AC conductivity \cite{strange} and fermions at finite density \cite{umut,liffermi,umut2}\footnote{Intriguingly, the  Lifshitz geometry appears to exhibit Ouroborosian stability properties, with  divergent tidal forces in the far infra-red leading to a stringy instability  \cite{way} which is sometimes resolved \cite{eva} and sometimes not, the latter case resulting in a background $AdS_2\times {\bf R}^2$ geometry \cite{sarah} which can itself be unstable to form a Lifshitz geometry \cite{strange} which has divergent tidal forces in the far infra-red leading to...}.

\para
Despite enjoying both energy and momentum conservation, when $z\neq 1$ certain processes exhibit a form of dissipation, even at zero temperature. This is seen most clearly in the  ballistic motion of objects --- whether lumps of energy created from the fields or massive probe particles --- both of which slow down, seemingly experiencing a friction force. There is no mystery here. Indeed, for relativistic $z=1$ theories the constant speed of the centre of mass is not due to momentum conservation but instead can be traced to invariance under boosts.  For $z\neq 1$, energy and momentum are conserved but drain into  the soft infra-red modes of the theory. The higher the value of $z$, the greater the number of these modes\footnote{This can be seen, for example, in the single particle density of states. A particle with dispersion relation $E \sim k^z$ in $d-1$ spatial dimensions has density of states $\rho(E)\sim E^{-1+(d-1)/z}$.} and the greater the friction.

\para
Moreover, for such ``dissipative" processes, there is sometimes a qualitative difference between  theories with $z<2$ and theories with $z>2$.  Perhaps the simplest example is the following: take a massive particle in the metric \eqn{metric} and throw it in a direction $x$ parallel to the boundary. Of course, the particle will follow a timelike geodesic, falling into the infra-red, $r\rightarrow 0$, of the geometry. The  question we would like to ask is: how far does the particle travel in the  $x$ direction? This is a simple exercise. One finds that for $z\leq 2$, the particle reaches $x\rightarrow \infty$ as $t\rightarrow \infty$. But for $z>2$, the particle only travels a finite distance.

\para
From the perspective of the boundary theory, this experiment corresponds to constructing a lump of  energy and giving it a kick. As the bulk particle falls towards $r\rightarrow 0$, the lump  disperses, while the motion in the $x$ direction tracks the centre of mass of the lump. For any $z> 1$, the lump slows down. But for $z>2$ the centre of mass never gets further than some finite distance\footnote{Although this statement holds true for any theory with a gravity dual, it does not appear to be a generic feature of quantum critical points. The counter-example is simply the non-relativistic Schr\"odinger equation, viewed as a classical field theory with $z=2$. The centre of mass travels with constant velocity.}.


\para
A zero temperature friction force also occurs for massive probe particles. In the context of string theory embeddings, these massive particles are modelled by strings attached to D-branes which sit at some fixed radial position in the background. The dynamics of such strings were first explored in AdS black hole backgrounds as a model of quarks ploughing through the quark-gluon plasma \cite{seattle}.  In Lifshitz spacetimes, analogous string dragging calculations reveal the existence of a friction force at zero temperature \cite{strange,kiritsis,mohsen,kazem}.  Furthermore, in the regime of linear response, there is again a difference between theories with $z<2$ and $z>2$. This manifests itself, for example, in  the AC conductivity due to massive charge carriers
which becomes non-analytic for $z>2$ \cite{strange}. (We will review the calculation behind this  in Section \ref{dissec}).

%
%

\para
There is a simple dimensional argument for the crossover in behaviour that occurs for $z=2$ \cite{strange}.   The kinetic term for a particle with inertial mass $m$ is
\be S = \frac{1}{2} \int dt\ m\, \dot{\vec{X}}\cdot\dot{\vec{X}}
\label{kinetic}\ee
The dimensions are inherited from the ambient critical fields, with scaling \eqn{scaling}. This means  $[X]=-1$ and $[t]=-z$. The mass then has dimension $[m]=2-z$ and the kinetic term on the worldline of the particle is irrelevant for $z>2$. The result is that the inertial mass, $m$, should play no role in the long-time dynamics of the particle when $z>2$. 

\para
The purpose of this short note is to elaborate on this $z=2$ crossover. Our main result, presented in the next section, is an analytic formula for the correlation function $\langle X(t) X(0)\rangle$ for a massive particle coupled to a quantum critical point. We will find that the long time behaviour is indeed independent of the inertial mass when  $z>2$. In Section 3, we revisit the calculation of the response function for the particle presented in \cite{strange}. We confirm the zero temperature fluctuation-dissipation theorem computed through the bulk variables. We further describe the diffusion of the particle at finite temperature.

\section{Fluctuation}

Throughout this paper we  model massive particles in the Lifshitz background as strings\footnote{Lifshitz geometries in supergravity  and, importantly, type II string theories have been constructed, starting with  \cite{strange,balas,jerome,ruth,varela}. The Lifshitz geometry also appeared in a different context in \cite{koroteev} and discussed in \cite{singh}.}, suspended from a D-brane that  fills the spacetime hypersurface at radial coordinate $r = r_b$. The string hangs from this D-brane, and extends down into the IR $r\rightarrow 0$ of the geometry \eqn{metric}.  While the endpoint of the string on the D-brane represents the position of the point particle, the string itself is to be interpreted as the strongly-coupled critical fields sourced by the particle.
  
\para
 In static gauge, the dynamics of the string are governed by the Nambu-Goto action
\be
S = -\frac{1}{2\pi \alpha '} \int_0^{r_b} dr dt\ r^{z-1} \sqrt{1+ r^4  \vec{x}^{\,\prime}\cdot \vec{x}^{\,\prime} - \frac{1}{r^{2z-2}}\dot{\vec{x}}\cdot  \dot{\vec{x}} }
\label{ng}\ee 
with Neumann boundary conditions imposed at the D-brane
\be
\vec{x}^{\,\prime}(r_b) = 0\label{neumann}\ee
The classical ground state of the string is simply the trivial solution $\vec{x}={\rm constant}$. The energy of this straight string is 
\be 
E = \frac{ 1}{2\pi \alpha'}\frac{r_b^z}{z}
\nn\ee
$E$ is the energy cost to create the string. For $z\neq 1$, this is not the same thing as the inertial mass $m$ in \eqn{kinetic}. Indeed, on dimensional grounds $[E]=z$ and $[m]=2-z$ so we expect 
\be m \sim E^{(2-z)/z}\label{eism}\ee
In Section \ref{dissec}, we will compute the response function and make this relation more precise. 

\para
Our goal is to understand the quantum fluctuations of the end point of the string,
\be \vec{X}(t)\equiv \vec{x}(t,r_b)\label{end}\ee
We do so by computing  the  two-point Green's function for the $x(t,r)$ fields, evaluated at $r_b$, namely $\langle X(t)X(0)\rangle$. For small fluctuations, we linearise the equations of motion about the average configuration $\vec{x}(t,r)=0$. The Nambu-Goto action reduces to 
\be
S \approx {\rm const} - \frac{1}{4\pi \alpha '} \int dr dt\, \left( r^{z+3}  \vec{x}^{\,\prime}\cdot\vec{x}^{\,\prime} - \frac 1 {r^{z-1}} \dot{\vec{x}}\cdot\dot{\vec{x}} \right)
\nn\ee
with the resulting equation of motion
\be
\frac{\partial}{\partial r}\left(r^{z+3} \frac{\partial \vec{x}}{\partial r} \right) - \frac{1}{r^{z-1}} \frac{\partial^2 \vec{x}}{\partial t^2} = 0\ .
\label{eom}\ee
Since the dynamics in each of the transverse dimensions decouple from one another, we will drop the vector notation and focus on just a single direction, $x$. 
%
%
%
Because the equation of motion is linear and invariant under time translations, we can Fourier decompose the most general solution as an integral over frequencies. 
\be
x(t,r) = \sqrt{\frac{\pi \alpha ' } z} \int_0^\infty d\omega\ U_\omega(r)\left[a(\omega) e^{-i\omega t} + a(\omega)^\dagger e^{i\omega t} \right]\label{sol1}\ee
where the modes are given by the $J$ and $Y$ Bessel functions in the combination
\be
U_\omega(r) = \frac{1}{\sqrt{1+C_\omega^2}}\frac{1}{r^{1+z/2}}\Big(J_{\frac 1 2 + \frac 1 z} 
 \left( \frac \omega {z r^z} \right) + C_\omega \, Y_{\frac 1 2 + \frac 1 z} \left( \frac \omega {z r^z} \right)  \Big)
 \label{mode}\ee
We have chosen a particular normalization of these modes which will be explained shortly. The coefficient $C_\omega$, which fixes the relative weights of the $J$ and $Y$ Bessel functions, is determined by the requirement of Neumann boundary conditions \eqn{neumann} at the D-brane. It is
\be
C_\omega = - \frac{ J_{\frac 1 z - \frac 1 2} \left( \frac \omega {z r_b^z} \right) } { Y_{\frac 1 z - \frac 1 2} \left( \frac \omega {z r_b^z} \right) }\ .
\nn\ee
At this stage, we turn to the quantum theory. The momentum conjugate to $x(t,r)$ is 
\be
\pi(t,r) = \frac 1 {2\pi \alpha' } \frac{1}{r^{z-1}}\, \dot x(t,r) \ ,
\nn\ee
and we impose the usual  equal-time canonical commutation relations
\be
[x(t,r), \pi(t,r')] = i\delta(r-r'), \qquad [x(t,r), x(t,r')] = [\pi(t,r), \pi(t,r')] = 0.
\nn\ee
The reason for the cleverly chosen normalization of \eqn{sol1} and \eqn{mode} now becomes apparent. Taking the inverse Fourier transform of \eqn{sol1}, and using various integral properties of the Bessel functions, we find that the equal time commutation relations above imply that the creation and annihilation operators are correctly normalized\footnote{In fact, there is more to this calculation than we're letting on. To do things carefully, we first impose an IR cut-off which renders the spectrum discrete, as in \cite{mukund}. This allows us to correctly normalize the modes, which then merge into a continuum as the cut-off is removed, resulting in \eqn{aacom}.}, 
\be
[a(\omega), a^\dagger(\omega')] = \delta (\omega - \omega'), \qquad [a(\omega), a(\omega')] = [a^\dagger (\omega), a^\dagger(\omega')] = 0.
\label{aacom}\ee
Finally, we define the vacuum state to be 
\be
a(\omega) \vert 0 \rangle = 0\ \ \ \ \forall\ \omega
\nn\ee 
This is the vacuum appropriate for observers that are stationary with respect to the  coordinates in \eqn{metric}.

\para
We now have everything in place to compute the two-point function of the end point of the string \eqn{end}. Substituting $r=r_b$ into the mode expansion \eqn{sol1} and engaging in some appropriate Besselology, we find
\be \langle X(t)X(0)\rangle = \int_0^\infty \frac{d\omega}{2\pi}\ e^{-i\omega t}\, \langle X(\omega) X(0)\rangle\label{cor1}\ee
with 
\be \langle X(\omega) X(0)\rangle =  8\alpha ' z \, \frac{r_b^{z-2}}{\omega^2} \left[J_{\frac 1 z - \frac 1 2}^2 \left( \frac{\omega}{z r^z_b} \right) + Y_{\frac 1 z - \frac 1 2}^2 \left( \frac{\omega}{z r_b^z} \right)\right]^{-1}
\label{result}\ee

It is instructive to look at the behaviour of the correlation function at low frequencies, for here we first see the crossover in behaviour at $z=2$ that we described in the introduction.  From the standard series expansions for the Bessel functions, we have that for $\omega \ll r_b^z$, 
\be
\langle X(\omega) X(0)\rangle\sim \left\{\ 
\begin{array}{cc} E^{2-\frac 4 z} \omega^{\frac 2 z - 3} &  \ \ \ \ \ \ \ \  1 < z< 2, \\
\omega^{-1-\frac 2 z} &  \ \ \ \ \ \ \ \  z > 2 \end{array}\right.
\label{lowf}\ee
Constant factors in these expressions have been suppressed for clarity. As advertised, the energy $E$ of the particle and, by \eqn{eism}, the inertial mass $m$, does not enter into the low frequency correlation for $z>2$.

\para
In the special case of $z = 1$ (pure AdS), the Fourier integral \eqn{cor1}  can be evaluated exactly to give
\be
\langle X(t) X(0)\rangle = -\frac 1 {4 \pi^2 \alpha 'E^2} (\log \vert t \vert + \gamma_E ) 
\nn\ee
For other values of $z$, no analytic expression for the inverse Fourier integral can be found. However, it is possible to obtain some estimate of the behaviour of the integrals at large times by assuming that the dominant contributions to the integrals come from the low frequency part of the integration range \eqn{lowf}. Using these expressions, we find the long time behaviour
\be
\langle X(t) X(0)\rangle\sim \left\{\ 
\begin{array}{cc} E^{2-\frac 4 z}  \vert t \vert^{2 - \frac 2 z}  &  \ \ \ \ \ \ \ \  1 < z< 2, \\
\vert t \vert ^{\frac 2 z} &  \ \ \ \ \ \ \ \  z > 2 \end{array}\right.
\nn\ee
A marked transition occurs at $z = 2$. Here the two-point function grows  linearly with $t$, which is its maximum rate of growth. The minimum growth occurs for $z=1$ and $z\rightarrow \infty$, both of which exhibit logarithmic behaviour. Moreover, for $z>2$, the long-time correlation of the particles has the peculiar behaviour of being independent of how heavy the particle is.

\section{Dissipation}\label{dissec}

We now turn to discuss the dissipation of momentum as the particle moves through the quantum critical bath. We will continue to work at zero temperature. However, at the end of this section we will briefly discuss the thermal contributions to the diffusion of the particle. 

\para
We start by computing the linear response of the particle to an external force $\vec{F}(\omega)$, 
\be \langle \vec{X}(\omega)\rangle = \chi(\omega) \vec{F}(\omega)\label{response}\ee
The essence of this calculation can be found in \cite{strange} (see also \cite{mukund}).  A force $\vec{F}(\omega)=\vec{E}e^{-i\omega t}$ is added to the endpoint of the string by the application of an oscillating electric field $E$ on the D-brane. The string action receives an additional boundary term from the associated gauge field $A_\mu$, 
\be
S = \left.\int dt \ A_t + \vec{A}\cdot\dot{\vec{x}}\ \right|_{r=r_b}
\nn\ee
Varying this action results in a new contribution to the boundary condition for the string endpoint. The Neumann boundary condition \eqn{neumann} is now replaced by
\be
 \vec{x}^{\,\prime}(r_b) = \frac{2\pi\alpha'}{r_b^{z+3}} \,\vec{F}.
\label{force}\ee
As in the previous section, we look for solutions to the equation of motion \eqn{eom}. We again restrict to motion in just a single direction, parallel to $\vec{F}$, now with  $x(t,r) = x(r) e^{-i\omega t}$. Solutions are again constructed of Bessel functions, but this time our boundary conditions differ, both on the D-brane and, crucially, in the infra-red. Let us start with the infra-red. Here we impose ingoing boundary conditions, appropriate to the computation of the retarded Green's function $\chi(\omega)$. These select the first Hankel function, $H^{(1)}=J+iY$,  over the second, 
\be x(t,r) = \frac{1}{r^{1+z/2}}\,H^{(1)}_{\frac{1}{2}+\frac{1}{z}}\left(\frac{\omega}{zr^z}\right)e^{-i\omega t}\nn\ee
This solution is supported by a force $F(\omega)$ which can be computed using \eqn{force}. It is
\be F(\omega) = \frac{1}{2\pi\alpha'}\,\omega  r_b^{1-z/2}\,H^{(1)}_{\frac{1}{z}-\frac{1}{2}}\left(\frac{\omega}{zr_b^z}\right)\nn\ee
Using \eqn{response}, we then have the response function
\be \chi(\omega) = \frac{2\pi\alpha'}{\omega r_b^2}\,\frac{H^{(1)}_{\frac{1}{z}+\frac{1}{2}}\left(\frac{\omega}{zr_b^z}\right)}{H^{(1)}_{\frac{1}{z}-\frac{1}{2}}\left(\frac{\omega}{zr_b^z}\right)}\label{responsible}\ee
It is again instructive to look at the low-frequency expansion of the response function.  For $\omega \ll r_b^z$, the expansion takes the form
\be \chi(\omega) \sim \frac{1}{m(i\omega)^2 + \gamma(-i\omega)^{1+2/z}+\ldots}\label{understand}\ee
with 
\be m = \frac{1}{(2-z)}\frac{1}{r_b^{z-2}}\ \ \ ,\ \ \ \gamma = \frac{1}{(2z)^{2/z}}\frac{\Gamma\left(\frac{1}{2}-\frac{1}{z}\right)}{\Gamma\left(\frac{1}{2}+\frac{1}{z}\right)}\nn\ee
The form of the response function \eqn{understand} has a very natural interpretation. $m$ is identified with the inertial mass \eqn{kinetic}. As we anticipated earlier, we see that it is indeed related to the energy as \eqn{eism}. However, perhaps more surprising, it changes sign at $z=2$. 

\para
The second term in the denominator of \eqn{understand} is the self-energy of the particle. For $z>2$, the self-energy dominates over the inertial mass at low frequencies. This is the manifestation of the simple dimensional analysis argument given in the introduction. It is responsible  for the non-analytic power-law behaviour in optical conductivity observed in \cite{strange}. Note that $\gamma$ also changes sign at $z=2$, and it does so in such a way that the ratio $\gamma/m$ is continuous at $z=2$ and the response function suffers no pathologies. 

\subsubsection*{The Fluctuation-Dissipation Theorem}

The fluctuation-dissipation theorem  relates the correlation function \eqn{result} to the imaginary, dissipative part of the response function
\be \langle X(\omega) X(0)\rangle = 2\left[  n_B(\omega)+1\right]\,{\rm Im}\,\chi(\omega)\label{flucdis}\ee
where $n_B(\omega) = (e^{\beta \omega}-1)^{-1}$ is the Bose-Einstein distribution and captures the dissipation due to thermal noise. The ``$+1$" in \eqn{flucdis} is the contribution due to quantum noise and in the zero temperature limit, $\beta\rightarrow\infty$, is all that survives. 

\para
Of course, it is straightforward to prove the fluctuation-dissipation theorem \eqn{flucdis} in quantum mechanics through spectral decomposition. Our purpose here is to verify this result from the bulk perspective. This has been previously studied in the AdS black hole background in \cite{mukund,son,teaney} where the black hole horizon and the associated finite temperature play a key role in the connection. There has also been related work on understanding current noise in the holographic context \cite{julian}. Here we verify that the result also holds for quantum dissipation at zero temperature in Lifshitz backgrounds. 

\para
Indeed, the result is straightforward. Using a standard Bessel function identity, the imaginary part of the full response function \eqn{responsible} is
\be {\rm Im}\,\chi(\omega) = \frac{4\alpha ' z r_b^{z-2}}{\omega^2}\,\frac{1}{J^2_{\frac{1}{z}-\frac{1}{2}}\left(\frac{\omega}{zr_b^z}\right)+Y^2_{\frac{1}{z}-\frac{1}{2}}\left(\frac{\omega}{zr_b^z}\right)}\nn\ee
As expected, this is proportional to the correlation function $\langle X(\omega) X(0)\rangle$ given in \eqn{result}. We have
\be {\rm Im} \,\chi(\omega) = \frac{1}{2}\langle X(\omega) X(0)\rangle, \nn\ee
confirming the zero temperature fluctuation-dissipation theorem for strings propagating in a Lifshitz background.

\subsection*{Thermal Diffusion}

For completeness, we finish with a discussion of dissipation at finite temperature.  In the present context, this means looking at the string correlation functions in the background of a Lifshitz black hole. There is now a rather extensive literature on constructing black holes in Lifshitz spacetimes. Many of these constructions are numerical, starting with \cite{lars}, while analytic solutions often involve a running dilaton breaking the scale invariance. (See, for example, \cite{stefan}).

\para
Here we give these difficulties short shrift. Following \cite{strange}, we simply work in a fixed background, paying no heed to the equations of motion that it solves.  The $(d+1)$-dimensional metric we choose is
\be
ds^2 = -r^{2z} f(r) dt^2 + \frac{dr^2}{f(r) r^2} + r^2 d\vec{x}\cdot d\vec{x}\nn
\ee 
The precise form of $f(r)$ will not concern us. We require only that $f(r)=0$ has a single zero at  the position of the horizon: $f(r)\sim (r-r_h)$ for $r\approx r_h$. The temperature of the boundary theory is the same as the Hawking temperature of the black hole, 
\be
 \beta = \frac {4\pi} {(z+d-1)}\frac{1}{r_h^z}.
\nn\ee
%
%
%
A massive particle in the hot Lifshitz bath is once again modelled by a string suspended from a D-brane placed at $r=r_b$. Our goal is to understand the thermal fluctuations of the endpoint of the string as it undergoes Brownian motion. This is a rather straightforward generalisation of the results in \cite{mukund,son,teaney} for asymptotically AdS black holes. Over long time scales, the endpoint is expected to wander diffusively, with the characteristic expectation value, now taken in the canonical ensemble, given by
\be \langle\, :\!(X(t)-X(0))^2\!:\,\rangle = 2(d-1) D t\label{thermalc}\ee
Here the $:$'s denote normal ordering. They remove the UV-divergent quantum fluctuations, leaving just the finite fluctuations due to the thermal bath. Our interest here is in computing the temperature dependence of the diffusion constant $D$. 

\para
Rather than compute the correlation function \eqn{thermalc} directly, we will instead invoke the fluctuation-dissipation theorem, now in the form of the Einstein relation. The diffusion constant can be related to the response function $\chi(\omega)$ defined in \eqn{response} by
\be
D = \frac {1} {\beta} \lim_{\omega \to 0}\left(-i\omega \chi(\omega)\right).
\label{einstein}\ee
Our goal, therefore, is once again to compute $\chi(\omega)$, now at finite temperature. The steps are exactly the same as the calculation that we just performed at zero temperature, but with one exception: we can't find exact solutions to the equations. The linearized equation of motion for the string in the black hole background is now
\be
 \frac{d }{dr}\left(r^{z+3} f(r) \frac{dx}{dr}\right)+\frac{\omega^2 }{r^{z-1}f(r)}x = 0.
\nn\ee
subject  to the boundary condition \eqn{force} on the D-brane. 
In terms of the tortoise radial coordinate $r^\star = \int dr f(r)^{-1} r^{-z-1}$, the equation of motion takes a Schrodinger-like form
\be
\left( \frac{d^2}{dr^{\star 2}} + \omega^2 - V(r) \right) (r x(r)) = 0,
\nn\ee 
where the effective potential is
\be
V(r) = r^{2z} f(r) \left( (z+1) f(r) + rf'(r)\right)
\nn\ee
Although we are unable to solve these equations analytically, all  that we require to determine the diffusion constant is the low frequency behaviour of the response function $\chi(\omega)$. Following \cite{mukund}, we calculate it by a patching procedure, splitting the spacetime into three regions:

\begin{itemize}
\item Region A: $r\sim r_h$, $V \ll \omega^2$
\item Region B: $r\sim r_h$, $V \gg \omega^2 $
\item Region C: $r \gg r_h$
\end{itemize}
In region A, we can drop the potential $V(r)$ from the equation of motion. We choose the causal, ingoing solution with $x_A = a e^{-i\omega r^\star}$ asymptotics. This is
\be
x_A(r) = a \left( 1 - \frac{i\omega}{(d+z-1) r_h^z} \log \left( \frac r {r_h} - 1 \right) + O(\omega^2) \right)
\nn\ee
In region C, the emblackening factor $f(r)$ is approximately unity, so the solution is the same superposition of Bessel functions \eqn{mode} that we found in the Lifshitz geometry at zero temperature. We are interested in the small $\omega$ behaviour of the solution, so we focus on the leading terms in its series expansion.
\be
x_C = c_1 \left( 1 +O(\omega^2) \right) + c_2 \left( \frac \omega {2z r^z} \right)^{1+ \frac 2 z} \left( 1 + O(\omega^2)\right)
\nn\ee
Finally, in region B, we drop the $\omega$ term from the equation of motion. The solution is
\be
x_B = b_1\left( 1 +O(\omega^2) \right) - b_2 r_h^{z+2} \int_r^\infty \frac{dr'}{{r'}^{z+3} f(r')}\left( 1 +O(\omega^2) \right).
\nn\ee
At the UV end of region B, $r \gg r_h$, we have $f(r)\approx 1$ and 
\be
x_B \approx b_1 \left( 1 +O(\omega^2) \right) - \frac{b_2 r_h^{z+2}}{(z+2) r^{z+2}} \left( 1 +O(\omega^2) \right),
\nn\ee
From this, we deduce that the solutions in regions B and C are consistent if
\be
b_1  = c_1, \qquad
b_2  =  - c_2 \frac{(z+2)}{r_h^{z+2}} \left( \frac \omega {2z} \right)^{1 + \frac 2 z}
\nn\ee 
At the IR end of region B, $r \sim r_h$, we need only use the fact that $f(r)$ has a single zero to find
\be
x_B \approx b_1 + b_2 \left( \frac{1}{d+z-1} \log \left( \frac{r}{r_h} - 1\right) + k \right).
\nn\ee
where $k$ is a constant of integration. Patching solutions in regions A and B requires
\be
b_1 + b_2 k = a, \qquad
b_2  = - \frac{i \omega a}{r_h^z}.
\nn\ee
Combining all of these results, we find that at lowest order in $\omega$,
\be
x_C(r_b) = a(1+O(\omega)), \qquad {x}^\prime_C(r_b) = \frac{i\omega r_h^2 a}{r_b^{z+3}}\,(1+O(\omega)).
\nn\ee
Thus we arrive at an expression for the response function at low frequencies
\be
 \chi(\omega)\approx \frac { 2\pi i \alpha '}{ r_h^2}\omega.
\nn\ee
Using the Einstein relation \eqn{einstein}, we learn what we set out to find: the dependence of the diffusion constant on the temperature is given by
\be
D \sim \beta^{\frac 2 z -1}.
\label{thisisd}\ee
Having struggled with us through our patching pain, the reader will be delighted to learn that this result can also be derived by some simple dimensional analysis. (A similar argument is given in \cite{mukund}).  If we assume that the mean free path of the particle depends only on the temperature, then $L_{\rm mfp}\sim T^{-1/z}$. Meanwhile, if the relaxation time similarly depends only on temperature then $\tau \sim 1/T$. In time $t$, the particle undergoes $N\sim t/\tau$ collisions and, assuming a diffusive mode of transport, travels  a distance $\Delta X=\sqrt{N}L_{\rm mfp}$. The upshot is
\be \Delta X^2 \sim \frac{L^2_{\rm mfp}}{\tau}\,t \sim Dt. \nn\ee
This reproduces \eqn{thisisd}.

\para 
In line with the general theme of this note, we find that there is again a difference as we cross $z=2$. For $z<2$, the rate of diffusion decreases with temperature. For $z>2$, we have the more usual situation where the rate increases with temperature. However, given the dimensional analysis argument above, this crossover does not seem to be related to the irrelevance of the inertial mass which occurs at zero temperature.


\section*{Acknowledgements}

We would like to thank Mike Blake, David Khmelnitskii, Mukund Rangamani, Julian Sonner and Andrei Starinets for useful discussions. We are supported by STFC and by the ERC STG grant 279943, ``Strongly Coupled Systems".

\end{document}